\documentclass[a4paper,11pt,fleqn]{article}
\pdfoutput=1 

\usepackage{jheppub} 

\usepackage[T1]{fontenc} 
\usepackage{slashed}

\usepackage{ucs}
\usepackage[utf8x]{inputenc}

\usepackage{amsmath}
\usepackage{amsfonts}
\usepackage{amssymb}
\usepackage[american]{babel}
\usepackage{array}
\usepackage{epsfig}
\usepackage{braket}
 \usepackage{feynmp}
 \usepackage{graphicx}
 \usepackage{feynmp}
 \usepackage{graphicx}
\newcommand{\Tr}{\text{tr~}}
\newcommand{\tr}{\text{tr~}}

\title{\boldmath On the Higher Spin Spectrum of Chern-Simons Theory coupled to Fermions in the Large Flavour Limit}

\author{V. Guru Charan,}
\author{and Shiroman Prakash}

\affiliation{Dept. of Physics and Computer Science, Dayalbagh Educational Institute, Agra, India 282005}

\emailAdd{vgurucharan2@gmail.com}
\emailAdd{shiroman@gmail.com}

\abstract{In this note, we compute the higher spin spectrum of $U(M)_k$ Chern-Simons theory coupled to $N$ flavours of fundamental fermions, in the limit $N\gg M$ with the 't Hooft coupling $\lambda_M =\frac{N}{k_m}$ held fixed, to order $M/N$. This theory possesses a slightly broken higher spin symmetry, and may be of interest from the perspective of higher-spin and non-supersymmetric holography. We find that anomalous dimensions of the higher spin currents achieve a finite value at strong coupling $\lambda_M \rightarrow \infty$, which grows with spin as $\log s$ for large $s$, as expected for gauge theories.}

\begin{document} 
\maketitle
\flushbottom

\section{Introduction and Summary of Results}
$U(M)_k$ Chern Simons theory coupled to $N$ flavours of fundamental fermions arises as a simple limit of $U(N)_{k_N}\times U(M)_{k_M}$ Chern-Simons theory coupled to bifundamental fermions. This theory is a generalization of the theory studied \cite{Giombi:2011kc}, and is of interest from the perspective of holography and higher-spin gauge theory, as we describe below.  The theory, and its cousin obtained by replacing the fermions with critical bosons has also been used in condensed matter physics as a calculable model to estimate critical exponents in fractional quantum Hall transitions (e.g., \cite{WenWu, Mott,Hui:2017pwe}). 

Theoretically, this theory is of interest because it is an example of a conformal field theory in $d>2$ for which explicit expressions for the scaling dimensions of higher-spin operators can be calculated at strong coupling, similar to vector models such as the critical $O(N)$ model, or the Gross-Neveu model in higher dimensions. (See \cite{Moshe:2003xn} for a review, and e.g., \cite{Muta:1976js, Lang:1992zw,Lang:1993ge, Giombi:2016hkj,Hikida:2016wqj,Nii:2016lpa,Hikida:2016cla, Manashov:2016uam, Giombi:2017rhm, Manashov:2017xtt} for calculations of scaling dimensions in vector models related to this paper) or the large flavour limit of 3d QED and QCD (e.g., \cite{Appelquist:1986fd, Appelquist:1986qw, Appelquist:1988sr, Appelquist:1989tc, PhysRevB.66.144501, Pufu:2013vpa}). With the revival (e.g.,\cite{Rattazzi:2008pe, Gopakumar:2016wkt, Dey:2016zbg}) of the bootstrap program \cite{Polyakov:1970xd}, various interesting methods, results and theorems about higher spin operators, e.g., \cite{Fitzpatrick:2012yx, Komargodski:2012ek, Alday:2013cwa, Kaviraj:2015cxa, Alday:2015eya, Kaviraj:2015xsa, Alday:2015ota, Skvortsov:2015pea} have been developed recently, and comparison to expectations from these analyses is another motivation for this work. It is possible to calculate S matrix exactly in these theories \cite{}.

\subsection{Motivation and Bifundamental Chern-Simons Theories}
As emphasized in \cite{Giombi:2011kc}, Chern-Simons theories coupled to matter are known to provide a large class of lines of non-supersymmetric conformal fixed points, in the large $N$ limit. Such lines are rare (or non-existent) in higher dimensions (e.g., \cite{Dymarsky:2005uh}) and are therefore of great interest. 

The specific example that we will focus our attention on in this paper is $U(N)_{k_1}\times U(M)_{k_2}$ Chern-Simons theory coupled to bifundamental fermions\cite{Gurucharan:2014cva}. This family of interacting theories is conformal in perturbation theory because the Chern-Simons levels $k_1$ and $k_2$ must be integers (or half-integers, depending on the number of fermion flavours), and therefore cannot run under RG flow. Because the scaling dimension of the fermion in three dimensions is $1$, there are simply no other marginal or relevant interactions one can write down (other than a mass term, which can easily be tuned to zero order-by-order in perturbation theory in any given renormalization scheme).

Let us assume $N\geq M$. Then a simple 't Hooft-like large $N$ limit that one can study is one in which $\alpha=\frac{M}{N}$, $\lambda_N=\frac{N}{k_N}$ and $\lambda_M=\frac{N}{k_M}$ are held fixed. We will be using a dimensional regularization scheme to define the Chern-Simons level $k$, so $|k_N|>N$ and $|k_M|>M$, which implies $-1<\lambda_N<1$ and $-\alpha^{-1}<\lambda_M<\alpha^{-1}$. 

In the 't Hooft limit described above, the parameters $\lambda_N$, $\lambda_M$ and $\alpha$ are effectively continuous, so we have a three-parameter family of non-supersymmetric interacting conformal field theories. Let us study the theory in an $M/N$ expansion, as discussed in \cite{Chang:2011mz} for the $\mathcal N=6$ ABJ theory \cite{Aharony:2008ug, Aharony:2008gk}. (See also \cite{Honda:2017nku} for a recent generalization of \cite{Chang:2012kt} to $\mathcal N=5$ supersymmetry.)

When $\alpha=M/N \ll 1$, the theory is essentially the vector model studied in \cite{Giombi:2011kc}, which is exactly solvable in the large $N$ limit. (See \cite{MZ, GurAri:2012is, Aharony:2012ns, Jain:2013py, Jain:2013gza, Jain:2014nza, Bedhotiya:2015uga, Gur-Ari:2015pca, Yokoyama:2016sbx, Minwalla:2015sca, Gur-Ari:2015pca, Inbasekar:2017ieo, Inbasekar:2017sqp, Inbasekar:2015tsa} for some all-orders results in Chern-Simons vector models.) The  single-trace primary operators of this theory consist of only a single tower of spin $s$ operators, one for each spin $s$. As argued in  \cite{Giombi:2011kc,Aharony:2011jz}, in this limit the anomalous dimensions of all single trace higher-spin primary operators vanish, i.e. $\tau_s = 1 + O\left(\frac{M}{N}\right)$. A bulk dual description must must therefore be a higher-spin gauge theory, and these operators would correspond to a tower of massless higher-spin gauge fields. Indeed, this higher-spin/vector-model duality \cite{Klebanov:2002ja, Sezgin:2002rt} has been well tested when $M=1$, \cite{Sleight:2016xqq, Sezgin:2003pt, Giombi:2009wh, Giombi:2010vg, Giombi:2011ya,  Giombi:2016ejx, Sezgin:2017jgm,Didenko:2017lsn} and we expect these results can be straightforwardly generalized to the bifundamental case in the limit $M/N \ll 1$, if one considers higher-spin gauge fields with $U(M)$ Chan Paton factors, as discussed for ABJ theory in \cite{Chang:2012kt}. 

When $M/N=1$, the theory is not a vector model, and instead can be thought of as a large $N$ theory with matrix degrees of freedom, completely analogous to ABJM theory \cite{Aharony:2008ug} or $\mathcal N=4$ SYM \cite{Maldacena:1997re}. Indeed, the theory can be thought of as a simple non-supersymmetric generalization of the ABJ(M) \cite{Aharony:2008ug, Aharony:2008gk} theory, that retains conformal invariance despite the lack of supersymmetry.

It is natural to ask whether the theory has a holographic dual at strong coupling when $M/N \sim 1$. We do not, at present, have a concrete way of testing any answer (see \cite{Gurucharan:2014cva} for suggestions) to this question since we are unable to perform any strong-coupling calculations when $M/N \sim 1$. However, as a preliminary step, we can calculate corrections to the scaling dimensions of higher-spin currents at strong coupling as a power series in $M/N$.

In this note we present some results for the first-order  $M/N$ corrections to the scaling dimensions of the single-trace, higher spin primaries which are twist one at zero coupling. The scaling dimensions depend on two parameters, $\lambda_M$ and $\lambda_N$, and take the following form, expressed in terms of the twist $\tau=\Delta-s$,:
\begin{equation}
    \tau_s=1 + \delta_s(\lambda_M,\lambda_N) \frac{M}{N}+O\left(\frac{M^2}{N^2}\right).
\end{equation}
If we set $\lambda_M=0$, then $\delta_s(0,\lambda_N)$ is simply $M$ times the $1/N$ anomalous dimension of $j_s$ computed in a Chern-Simons vector model in \cite{Giombi:2016zwa}, which we reproduce here: 
\begin{equation}
\delta_s(0,\lambda_N) = \frac{\pi \lambda_N}{2 \sin(\pi \lambda_N)}\left(a_s  \sin^2\left(\frac{\pi\lambda_N}{2}\right) +\frac{b_s}{4} \sin^2\left(\pi\lambda_N\right)\right)\,. \label{general-structure}
\end{equation}
with
\begin{eqnarray}
	a_s & = & \begin{cases} \frac{16}{3\pi^2} \frac{s-2}{2s-1}\,, & \text{for even $s$}\,, \\
 \frac{32}{3\pi^2} \frac{s^2-1}{4s^2-1}\,, &  \text{ for odd $s$}\,,\end{cases} \\
	\label{bs-eq-intro}
 b_s & = & \begin{cases}
 \frac{2}{3 \pi
   ^2} \left(3 (H_{s-\frac{1}{2}}+2\log (2))+\frac{-38 s^4+24 s^3+34 s^2-24
   s-32}{4
   s^4-5 s^2+1}\right)\,, & \text{for even $s$}\,, \\
    \frac{2}{3 \pi ^2} \left(3
   (H_{s-\frac{1}{2}}+2\log (2))+\frac{20-38 s^2}{4
   s^2-1}\right)\,, & \text{ for odd $s$}\,.
\end{cases}
\end{eqnarray}
The form \eqref{general-structure} follows from the analysis based on slightly-broken higher spin symmetry given in \cite{Maldacena:2011jn,Maldacena:2012sf}, and the spin-dependent numerical coefficients were determined using techniques developed in \cite{Giombi:2016hkj,Skvortsov:2015pea} as well as a direct two-loop Feynman diagram computation. When $s$ is large, $a_s \sim \frac{8}{3\pi^2}$ and $b_s \simeq \frac{2}{\pi^2}\log s$. Note that, as mentioned above, $|\lambda_N|<1$, so the logarthmic growth of anomalous dimension on spin is only present for intermediate values of the coupling. 

In this paper, we consider the higher-spin spectrum in a different limit, namely $\lambda_N=0$ and calculate the first-order $M/N$ spectrum of higher spin operators to all orders in $\lambda_M$. This limit can be essentially thought of as a large flavour limit, with $N$ playing the role of a large number of flavours. (However, the flavour symmetry is gauged, so the only gauge-invariant operators are singlets of the flavour symmetry). Below, we refer to the theory with $\lambda_N=0$ as the ``large flavour" theory, and the theory with $\lambda_M=0$ as the ``large colour" theory. 

\subsection{Summary of Results}

Our results are as follows:
\begin{equation}
\begin{split}
\delta_s(\lambda_M,0) = &  \frac{16}{3 \left(\pi ^2 \lambda_M ^2+64\right) }  {\left(3 H_{s-\frac{1}{2}}+6\log (2)-\frac{2 \left(11 s^4+3 s^3-13 s^2+15 s+2\right)}{4
   s^4-5 s^2+1}\right)}\\ &+\frac{64^2}{\left(\pi ^2 \lambda_M ^2+64\right)^2} \frac{
   (s-2) (s-1)^2}{  2\left(4 s^4-5 s^2+1\right)},  \text{ for even $s$},
   \end{split}
 \end{equation}
and 
\begin{equation}
    \delta_s(\lambda_M,0) = 
 \frac{16 \lambda_M^2  \left(3 H_{s-\frac{1}{2}}+6\log
   (2)+\frac{4-22 s^2}{4 s^2-1}\right)}{3 \left(\pi ^2 \lambda_M ^2+64\right)},   \text{ for odd $s$}.
\end{equation}
Here $H_s$ are harmonic numbers. In section \ref{hs}, we argue that the $\lambda_M$-dependence of these anomalous dimensions follows from the planar three-point functions and the higher spin symmetry, up to spin dependent coefficients. We then fix these spin dependent coefficients in the direct diagrammatic calculation presented in section \ref{diagram-section}. 

The anomalous dimensions of higher-spin currents in this limit acquire finite values as $\lambda_M \rightarrow \infty$,  which increase with spin as $\log s$ for large spin:
\begin{equation}
  \delta_s(\lambda_M,0) \sim  \frac{16 \lambda_M ^2}{\left(\pi ^2 \lambda_M
   ^2+64\right)} \log s\,.
\end{equation}
The logarithmic dependence on spin is a characteristic feature of gauge theory that is expected from general arguments \cite{Alday:2007mf}. In section \ref{sec-3} we present the large spin expansion of this spectrum in a bit more detail, and discuss to what extent these results follow from a more general analysis of expectations for the large-spin spectrum of conformal field theories given in \cite{Alday:2015eya, Alday:2015ota}.

\section{$M/N$ Gauge-Propagator}

See \cite{Gurucharan:2014cva} for detailed description of the theory we study and our conventions.

Here we are working in the limit $\lambda_N=0$, and calculating the $M/N$ correction to the anomalous dimension to all orders in $\lambda_M$. In this limit, the gauge propagator is suppressed by a factor of $1/N$, but it receives an infinite series of self-energy corrections to first order in $M/N$, which are given by: 

\begin{fmffile}{gaugex}
        \begin{tabular}{c}
            \begin{fmfgraph*}(60,40)
                \fmfleft{i1}
                \fmfright{o1}
 \fmfv{decoration.shape=circle, decoration.size=.2h, decoration.filled=shaded}{z}
		 \fmf{photon,tension=.5}{i1,z,o1}
             \end{fmfgraph*} 
        \end{tabular}
 \end{fmffile}
	$=$ 
\begin{fmffile}{gauge0}
        \begin{tabular}{c}
            \begin{fmfgraph*}(40,40)
                \fmfleft{i1}
                \fmfright{o1}
 		 \fmf{photon,tension=5}{i1,o1}
             \end{fmfgraph*}
        \end{tabular}
        \end{fmffile}
$+$
\begin{fmffile}{gauge1}
        \begin{tabular}{c}
            \begin{fmfgraph*}(60,40)
                \fmfleft{i1}
                \fmfright{o1}
		\fmf{fermion,left, tension=2}{a,b,a}
		\fmf{photon,tension=5}{i1,a}
		\fmf{photon, tension=5}{b,o1}
             \end{fmfgraph*}
        \end{tabular}
        \end{fmffile}
$+$
\begin{fmffile}{gauge2}
        \begin{tabular}{c}
            \begin{fmfgraph*}(80,40)
                \fmfleft{i1}
                \fmfright{o1}
		\fmf{fermion,left, tension=1.5}{a,b,a}
	        \fmf{fermion,left, tension=1.5}{c,d,c}
		\fmf{photon,tension=5}{i1,a}
		\fmf{photon, tension=5}{b,c}
		\fmf{photon,tension=5}{d,o1}
             \end{fmfgraph*}
\end{tabular}
$ + ~ \ldots $.
\end{fmffile}

The one-loop self energy of the gauge field, $\Sigma^{\mu\nu}$ is given by 
\begin{equation}
 \Sigma^{\mu\nu}(q)  = -\frac{N}{2} (-1)\tr \int \frac{d^3p}{(2\pi)^3}
\frac{1}{i p_\alpha \gamma^\alpha} \gamma^\mu \frac{1}{i
(p_\beta+q_\beta) \gamma^\beta} \gamma^\nu = -N \frac{q^2 \delta^{\mu\nu} - q^\mu
q^\nu}{32 q}
\end{equation}

Let $D_{\mu\nu}$ be the free gauge propagator and let $G_{\mu\nu}$ be the gauge
propagator with an infinite series of self-energy corrections. Then, the matrices $\mathbf{G}$ and $\mathbf{D}$ are related by
$
 \mathbf{G} = \left( 1 - \mathbf{D}\mathbf{\Sigma} \right)^{-1} \mathbf{D}
$.

In light-cone gauge, which we will use throughout the paper, the corrected gauge propagator is
\begin{equation}
\mathbf{G} = \begin{pmatrix} G_{33} & G_{3+} \\ G_{+3} & G_{++}
              \end{pmatrix}
= \frac{1}{N} \frac{2\pi^2q_+^2}{q q_s^4} \frac{64}{64+\pi^2 \lambda_M^2}
\begin{pmatrix} \lambda_M^2 q_-^2  & \frac{8i\lambda_M}{\pi} q_- q - \lambda_M^2 q_-
q_3 \\
 -\frac{8i\lambda_M}{\pi} q_- q - \lambda_M^2 q_- q_3  & - q_s^2 \lambda_M^2
 \end{pmatrix}
\end{equation}
Explicitly,
\begin{eqnarray}
G_{33} & = & A \frac{\pi^2}{2N} \frac{\lambda_M^2}{q} \\
G_{3+} &= & A\frac{8 i \pi \lambda_M}{N} \frac{q_+ }{q_s^2} - A\frac{\pi^2 \lambda_M^2}{N} \frac{q_3 q_+}{q q_s^2} \\
G_{+3} &=& A\frac{-8 i \pi \lambda_M}{N} \frac{q_+ }{q_s^2} - A\frac{\pi^2 \lambda_M^2}{N} \frac{q_3 q_+}{q q_s^2} \\
G_{++} &=& A\frac{- 2\pi^2 \lambda_M^2}{N} \frac{q_+^2}{q q_s^2}
\end{eqnarray}
where 
\begin{equation}
    A= 64(64+\pi^2\lambda_M^2)^{-1}.
\end{equation}

\section{Analysis based on slightly-broken higher-spin symmetry}
\label{hs}

\subsection{Higher Spin Currents}
The higher-spin currents $j_s(x,z)$ are given by the following explicit expressions \cite{Giombi:2011kc,Giombi:2016zwa}:
\begin{equation}
j_s^F(x,z) = f_s^F(\partial_{x_1}\cdot z, \partial_{x_2}\cdot z) \bar{\psi}(x_1) \slashed{z} \psi(x_2)\Big|_{x_1=x_2=x}
\end{equation}
where
\begin{equation}
    f_s^F(u,v)=\sum_{k=0}^{s-1} \frac{(-1)^{k+s+1}}{2s!} \binom{2s}{2k+1} u^k v^{s-k-1}.
\end{equation}
Here, and in what follows, $z$ is a null polarization vector satisfying $z^2=0$ (see e.g., \cite{Giombi:2016hkj, Costa:2011mg}), and we are suppressing color-indices. These expressions are defined for the free theory. In the interacting theory, derivatives are promoted to covariant derivatives.

It was argued in \cite{Giombi:2011kc} that these operators, along with the scalar primary $j_0=\bar{\psi}\psi$, are the only single-trace primary operators in the Chern-Simons vector model. In the bifundamental theory, this remains true, if we define ``single-trace" operators to mean bilinear operators constructed from a single contraction of the $U(N)$ color index, which is natural in the limit $N$ large and $N \gg M$ \cite{Chang:2012kt}.

In momentum space, these operators can be written as:
\begin{equation}
j_s(q,z) = \int \frac{d^3p}{(2\pi)^3}  \bar{\psi}(q-p)f_s^F(i(q-p)\cdot z,ip\cdot z) \slashed{z}\psi(p).
\end{equation}
We define the free vertex as:
\begin{eqnarray}
V_s^0(q,p) & = &  \gamma^\mu z_\mu f_s^F(i(q-p)\cdot z,ip\cdot z).
\end{eqnarray}
When $q=0$, this simplifies to
\begin{eqnarray}
V_s^0(0,p) & = & \slashed{z} {4^s \over 2s!} (-ip\cdot z)^{s-1} \\
& = & V_s~(p \cdot z)^{s-1} \slashed{z}.
\end{eqnarray}

The anomalous dimension, $\delta_s=\tau_s-1$, of $j_s$ is related to the logarithmic divergence of the corrected vertex $V'(q,p)$ via $V'_s(0,p) =-\delta_s V^0_s(0,p) \log \Lambda$.

\subsection{General Form of the Anomalous Dimensions}

Let us first derive expressions for the general form of the anomalous dimensions using the slightly-broken higher spin symmetry of the theory. As argued in \cite{MZ, Giombi:2016zwa}, we can determine the $1/N$ anomalous dimension of $j_s$ from the leading order (planar) parity-violating three-point functions $\langle j_{s_1} j_{s_2} j_{s} \rangle$, outside the `triangle inequality', (i.e., with $s_1+s_2<s$) \cite{Giombi:2011rz,MZ}. 

Let us review this briefly. For simplicity, let us take $M=1$ in what follows (although all these expressions can be generalized to $M \neq 0$ as long as $M \ll N$). 
\begin{equation}
    \partial \cdot j_s = \sum_{s_1,s_2}\frac{1}{N}C_{s_1,s_2,s}(\lambda_M) [j_{s_1}] [j_{s_2}], \label{anom-div}
\end{equation}
where $[j_{s_1}] [j_{s_2}]$ denotes the particular combination of descendents of $j_{s_1}$ and $ j_{s_2}$ given in \cite{Giombi:2016zwa}. Note that the exact form of $[j_{s_1}][j_{s_2}]$ is uniquely determined by the requirement that $\partial \cdot j_s$ is a conformal primary. Also, note that the only spins that can appear on the right hand side of \eqref{anom-div} are those satisfying $s_1+s_2<s$.

Then, via the state operator correspondence, (see, e.g., \cite{Giombi:2011kc,MZ}) 
\begin{equation}
    \braket{\partial \cdot j_s|\partial \cdot j_s} \sim (\tau_s-1) \langle j_s | j_s \rangle.
\end{equation}
This implies that, 
\begin{equation}
    \tau_s-1 = \sum_{s_1,s_2} \frac{C_{s_1,s_2,s}^2}{N^2} \alpha_{s_1,s_2,s}\frac{\langle j_{s_1} j_{s_1} \rangle \langle j_{s_2} j_{s_2} \rangle }{\langle j_s j_s \rangle}, \label{twist-eqn} 
\end{equation}
where $\alpha_{s_1,s_2,s}$ is a purely numerical coefficient that can be computed using the particular combination of descendents of $j_{s_1}$ and $j_{s_2}$ represented by $[j_{s_1}][j_{s_2}]$. 

We can determine the $C_{s_1, s_2, s}(\lambda_M)$ from the planar three-point functions as follows
\begin{equation}
    \langle j_{s_1} j_{s_2} \partial \cdot j_s \rangle \sim \frac{1}{N}C_{s_1,s_2,s}(\lambda_M) \langle j_{s_1} j_{s_1} \rangle \langle j_{s_2} j_{s_2} \rangle,
\end{equation} which implies
\begin{equation}
    C_{s_1,s_2,s} \sim N \frac{\langle j_{s_1} j_{s_2} j_s \rangle}{\langle j_{s_1} j_{s_1} \rangle \langle j_{s_2} j_{s_2} \rangle}.
\end{equation}
Here $\sim$ denotes equality up to a numerical coefficient that again depends on the details of the form of $[j_{s_1}][j_{s_2}]$.

Although this is a theory with a slightly-broken higher spin symmetry its planar three-point functions have not been computed in \cite{MZ}, which applies only to theories containing only even spins (e.g., $O(N)$ theories with Majorana fermions). However, it is straightforward to compute them in momentum space directly. Loops of the gauge field are suppressed by $1/N$ in the large flavour limit, so we only consider tree-level diagrams. It is easy to see that nearly all three point functions are those of the theory of free fermions -- the only exceptions are three-point functions involving the spin $1$ current $j_1$, pictured in figure \ref{3point-1} and \ref{3point-2}. The $U(M)$ Chern-Simons gauge field effectively acts as a double trace deformation of the schematic form $j_1 j_1$ in the action, once integrated out.  
\begin{figure}[h]
\center
\begin{fmffile}{threepointfunctionone}
        \begin{fmfgraph*}(200,160)
                \fmfleft{ii1,ii2}
		\fmfright{oo1,oo2}
		\fmffixed{(.05w,.5h)}{ii1,i1}
	    \fmffixed{(.05w,-.5h)}{ii2,i1}
        \fmffixed{(.3w,.2h)}{i1,x}
		\fmffixed{(.05w,-.05h)}{o1,oo1}
		
		\fmffixed{(0,-.4h)}{x,y}
		\fmflabel{$j_{s'}$}{x}
		\fmflabel{$j_1$}{o1}
		\fmflabel{$j_{s}$}{i1}
		  \fmfv{decoration.shape=circle, decoration.size=.1h, decoration.filled=shaded}{b}
		\fmf{photon,tension=.5}{y,b,z}
        \fmf{electron,  tension=.3}{i1,x,y,i1}
                \fmf{electron, left, tension=.3}{o1,z,o1}
		\fmfv{decoration.shape=circle, decoration.size=.025h}{i1}
		\fmfv{decoration.shape=circle, decoration.size=.025h}{x}
		\fmfv{decoration.shape=circle, decoration.size=.025h}{o1}
             \end{fmfgraph*}
    \end{fmffile}
    \caption{Diagram contributing to the three point function $\langle j_1 j_{s'} j_s\rangle$. \label{3point-1}}
\end{figure}
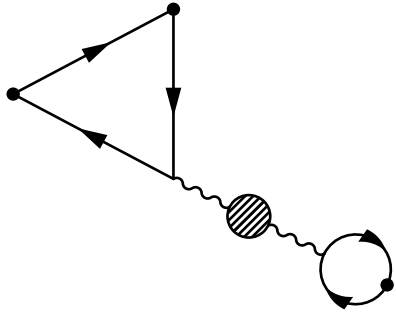
 \begin{figure}[h]
\center
\vspace{.5cm}
\begin{fmffile}{threepointfunction2}
        \begin{fmfgraph*}(200,160)
                \fmfleft{ii1,ii2}
		\fmfright{oo1,oo2}
		\fmffixed{(.05w,.5h)}{ii1,i1}
	    \fmffixed{(.05w,-.5h)}{ii2,i1}
        \fmffixed{(.3w,.2h)}{i1,x}
		\fmffixed{(.05w,-0.09h)}{o1,oo1}
		\fmffixed{(.05w,+.09h)}{o2,oo2}

		\fmffixed{(0,-.4h)}{x,y}
		\fmflabel{$j_1$}{o2}
		\fmflabel{$j_1$}{o1}
		\fmflabel{$j_{s}$}{i1}
		  \fmfv{decoration.shape=circle, decoration.size=.1h, decoration.filled=shaded}{b}
		\fmf{photon,tension=.5}{y,b,z}
        \fmf{electron,  tension=.3}{i1,x,y,i1}
            \fmf{electron, left, tension=.3}{o1,z,o1}
		\fmfv{decoration.shape=circle, decoration.size=.025h}{i1}
        \fmfv{decoration.shape=circle, decoration.size=.1h, decoration.filled=shaded}{a}
		\fmfv{decoration.shape=circle, decoration.size=.025h}{o1}
		\fmf{photon,tension=.5}{x,a,w}
		\fmfv{decoration.shape=circle, decoration.size=.025h}{o2}
		\fmf{electron, left, tension=.3}{o2,w,o2}
             \end{fmfgraph*}
    \end{fmffile}
    \caption{Diagram contributing to the three point function $\langle j_{1} j_1 j_s \rangle$. \label{3point-2}}
\end{figure}
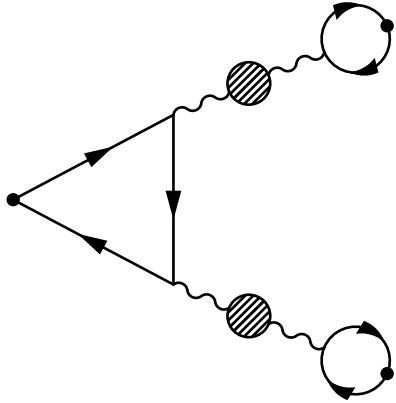

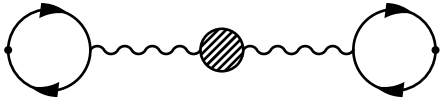
\begin{figure}[h]
\center
\begin{fmffile}{twopointfunction}
        \begin{fmfgraph*}(200,80)
                \fmfleft{ii1,ii2}
		\fmfright{oo1,oo2}
		\fmffixed{(.05w,.5h)}{ii1,i1}
	    \fmffixed{(.05w,-.5h)}{ii2,i1}
        \fmffixed{(.3w,.2h)}{i1,x}
		\fmffixed{(.05w,0.5h)}{oo1,o1}
		 \fmffixed{(.05w,-.5h)}{oo2,o1}
		\fmflabel{$j_1$}{o1}
		\fmflabel{$j_1$}{i1}
		  \fmfv{decoration.shape=circle, decoration.size=.2h, decoration.filled=shaded}{b}
		\fmf{photon,tension=.5}{y,b,z}
      \fmf{electron, left, tension=.4}{i1,y,i1}
\fmf{electron, left, tension=.4}{o1,z,o1}
		\fmfv{decoration.shape=circle, decoration.size=.025h}{i1}
		\fmfv{decoration.shape=circle, decoration.size=.025h}{o1}
             \end{fmfgraph*}
    \end{fmffile}
    \caption{Diagram contributing to $\langle j_1 j_1 \rangle$ \label{two-point}}
\end{figure}

The non-vanishing parity-odd three point functions, outside the triangle inequality, are only $\langle j_1 j_{s'} j_s \rangle$ and $\langle j_1 j_1 j_s \rangle$, where $s,s'\neq 1$, pictured in Figures \ref{3point-1} and \ref{3point-2}. The second three-point function is only non-zero when $s$ is even \cite{Giombi:2011kc}. Using the the corrected gauge propagator, we can directly read off the $\lambda_M$-dependence of these three point functions:
\begin{eqnarray}
\langle j_1 j_{s'} j_s \rangle\Big|_{\text{parity odd}} & \sim & (\mathbf 1-\mathbf{D\Sigma})^{-1} \Big|_{\text{parity odd}} \sim  A\lambda_M N\\
\langle j_1 j_{1} j_s \rangle\Big|_{\text{parity odd}} & \sim & A^2\lambda_M N 
\end{eqnarray}
We normalize two-point functions as $\langle j_{s} j_{s}\rangle \sim N$, but the two-point function of $j_1$ is given by
\begin{equation}
    \langle j_{1} j_1 \rangle \sim \mathbf \Sigma + \mathbf{\Sigma D \Sigma}+\dots = \mathbf {\Sigma (1-D\Sigma)}^{-1} \sim A N,
\end{equation}
as can be seen from Figure \ref{two-point}.
This implies that
\begin{eqnarray}
C_{1,s',s} & \sim & {\lambda_M} \\
C_{1,1,s} & \sim & {\lambda_M}.
\end{eqnarray}

The anomalous dimensions are then given by:
\begin{equation}
    \tau_s-1 = (c^{(1)}_s A \lambda_M^2 + c^{(2)}_s A^2\lambda_M^2) \frac{1}{N}.\label{general-form}
\end{equation}
where $c^{(1)}_s$ and $c^{(2)}_s$ are spin-dependent numerical coefficients, and $c^{(2)}_s$ is non-zero only if $s$ is even because $\langle j_1 j_1 j_s \rangle$ vanishes when $s$ is odd. When $M \neq 1$, one can show that equation \eqref{general-form} is multiplied by $M$, as long as $M \ll N$. (However, this is not true at higher orders, as $M^2/N^2$ corrections will differ from $1/N^2$ corrections.)

In the section below we directly evaluate the logarithmic divergences of the relevant Feynman diagrams to compute the anomalous dimension and verify that it is of the form \eqref{general-form}. From the calculation we are also able to read off the coefficients $c^{(1)}$ and $c^{(2)}$, and find $c^{(2)}=0$ when $s$ is odd, as expected. The results are:
\begin{eqnarray}
    c^{(1)} & = & \begin{cases}
    \frac{1}{12} {\left(3 H_{s-\frac{1}{2}}+6\log
   (2)+\frac{4-22 s^2}{4 s^2-1}\right)} & \text{ for $s$ odd,} \\
  \frac{1}{12}{\left(3 H_{s-\frac{1}{2}}+6 \log (2)-\frac{2 \left(11 s^4+3 s^3-13 s^2+15 s+2\right)}{4
   s^4-5 s^2+1}\right)} & \text{ for $s$ even,}
   \end{cases} \label{spin-1}\\
    c^{(2)} & = &
    \begin{cases}
    0 & \text{ for $s$ odd,} \\
     \frac{ (s-2) (s-1)^2}{2  \left(4 s^4-5 s^2+1\right)}
     &  \text{ for $s$ even.}
     \end{cases} \label{spin-2}
\end{eqnarray}

We conclude by noting that one could also use the classical equations of motion to determine $C_{s_1,s_2,s}$ to order $\lambda_M^2$. The coefficients $C_{s_1,s_2,s}$ in this large-flavour theory are a priori expected to be different from those in the large-colour theory \cite{Giombi:2016zwa}, since the theories are different. For example, for $s_1,s_2\neq 1$, $C_{s_1,s_2,s}$ are zero for the large flavour theory but non-zero for the large colour theory. Then $c^{(2)}$ would be related to $C_{1,1,s}^2$, and $c^{(1)}$ would be related to the $C_{1,s',s}^2$ via equations very similar to equations (3.27)-(3.29) of \cite{Giombi:2016zwa}.

\section{Direct Feynman Diagram Calculation}
\label{diagram-section}
The calculation of the anomalous dimension of the scalar $\bar{\psi}\psi$ was carried out in \cite{Gurucharan:2014cva}. Here, we consider the anomalous dimensions for currents with $s>0$ which follows  \cite{Gurucharan:2014cva}. 

The anomalous dimension is given by 
\begin{equation}
    \delta_s=\delta^{(1)}_s+\delta^{(2)}_s+\delta^{(3)}_s
\end{equation}
where $\delta^{(1)}_s$ is the contribution from the self-energy diagram, $\delta^{(2)}_s$ is the contribution from the rainbow diagram, $\delta^{(3)}_s$ is the contribution from the two three-point functions. 

\subsection{Self-Energy}

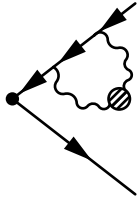
\begin{figure}[h]
\begin{center}
\begin{fmffile}{Vertex1}
        \begin{tabular}{c}
            \begin{fmfgraph*}(60,80)
                \fmfleft{i1}
                \fmfright{o1,o2}
                \fmffixed{(-.1h,0)}{y1,i1}
                \fmffixed{(0,-.05h)}{y2,o1}
                \fmffixed{(0,.05h)}{y3,o2}
                \fmfv{decoration.shape=circle, decoration.size=.05h}{y1}
                \fmffixed{(0,.5h)}{o1,v}
                \fmffixed{(0,.5h)}{v,o2}
                \fmffixed{(-.5h,0)}{z,y1}
                \fmf{photon, left=.6,tension=.0}{x,z,y}
                \fmfv{decoration.shape=circle, decoration.size=.1h,
decoration.filled=shaded}{z}
                \fmf{electron}{y3,x,y,y1,y2}
                \fmflabel{$j_s$}{y1}
             \end{fmfgraph*}
        \end{tabular}
        \end{fmffile}
\caption{Self energy correction to the vertex}
\end{center}
\end{figure}

The contribution from the fermion self energy to the logarithmic divergence of the two-point function $\langle j_s j_s \rangle$ is
\begin{equation}
    \frac{\left( -\frac{1}{6} I_A + I_B + 2 I_C \right)}{I_A} A \frac{\lambda_M^2 M}{N} \frac{1}{4} \log \Lambda = \delta_1 (-\log \Lambda)
\end{equation}
where
\begin{eqnarray}
I_A & = &  - \int \frac{d^3p}{(2\pi)^3} \tr \left( \frac{1}{i\slashed{p}} \gamma_- \frac{1}{i(\slashed{p}-\slashed{q})} \gamma_- \right) q_-^{s-1} (q_--p_-)^{s-1} \\
I_B & = & - \int \frac{d^3p}{(2\pi)^3} \tr \left(\frac{1}{i\slashed{p}} i p_3 \gamma_3 \frac{1}{i\slashed{p}} \gamma_- \frac{1}{i(\slashed{p}-\slashed{q})} \gamma_- \right) q_-^{s-1} (q_--p_-)^{s-1} \\
I_C & = & - \int \frac{d^3p}{(2\pi)^3} \tr \left(\frac{1}{i\slashed{p}} i p_+ \gamma_- \frac{1}{i\slashed{p}} \gamma_- \frac{1}{i(\slashed{p}-\slashed{q})} \gamma_- \right) q_-^{s-1} (q_--p_-)^{s-1} 
\end{eqnarray}
We find the contribution of self-energy correction to anomalous dimensions is: 
\begin{equation}
  \delta^{(1)}_s= -\frac{11}{24} A\lambda_M^2 \frac{M}{N}
\end{equation}
\vspace{1mm}
\subsection{Rainbow Correction}
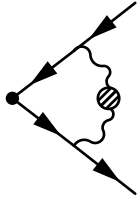
\begin{figure}[h]
\begin{center}
 \begin{fmffile}{Vertex2}
        \begin{tabular}{c}
            \begin{fmfgraph*}(60,80)
                \fmfleft{i1}
                \fmfright{o1,o2}
                \fmffixed{(-.1h,0)}{y1,i1}
                \fmffixed{(0,-.05h)}{y2,o1}
                \fmffixed{(0,.05h)}{y3,o2}
               \fmfv{decoration.shape=circle, decoration.size=.05h}{y1}
                \fmffixed{(0,.5h)}{o1,v}
                \fmffixed{(-.45h,0)}{z,y1}
                \fmffixed{(0,.5h)}{v,o2}
                \fmf{photon,right=.5,tension=.01}{x,z,y}
                \fmfv{decoration.shape=circle, decoration.size=.1h,
decoration.filled=shaded}{z}
                \fmf{electron}{y3,y,y1,x,y2}
                \fmflabel{$j_s$}{y1}
             \end{fmfgraph*}
        \end{tabular}
        \end{fmffile}
\caption{The rainbow correction to the vertex.}
\end{center}
\end{figure}
The rainbow correction is determined by the following integral
\begin{equation}
    I_2 = V_s \frac{M}{2}  \frac{1}{2} \Tr \left(\gamma^- \int \frac{d^3q}{(2\pi)^3} \left(G_{\mu\nu}(q) \gamma^\mu \frac{1}{i (\slashed{p}+\slashed{q})} \gamma_- (p_-+q_-)^{s-1}  \frac{1}{i (\slashed{p}+\slashed{q})} \gamma^\nu \right) \right) 
\end{equation}
which contributes to the logarithmic divergence via
\begin{equation}
    I_2 = V_s p_-^{s-1} \delta_2 (-\log \Lambda)
\end{equation}
and we find:
\begin{equation}
      \delta^{(2)}_s =  \frac{1}{4}  \left(- \frac{1}{2(4s^2-1)}+g(s) \right) \lambda_M^2 A \frac{M }{N}
\end{equation}
where, $g(s)$ is
\begin{equation}
  g(s)=  \gamma -\psi(s)+2\psi(2s) =  \sum_{n=1}^s \frac{1}{n-1/2} = H_{s-\frac{1}{2}}+\log (4).
\end{equation}

\subsection{Other Contributions}

\begin{figure}[h]
\center
\begin{tabular}{cc}
\begin{fmffile}{Additional}
            \begin{fmfgraph*}(120,80)
                \fmfleft{ii1}
		\fmfright{oo1,oo2}
		\fmffixed{(.05w,0)}{ii1,i1}
		\fmffixed{(.05w,0)}{o1,oo1}
		\fmffixed{(.05w,0)}{o2,oo2}
		\fmffixed{(0,.5h)}{o1,v}
		\fmffixed{(0,.5h)}{v,o2}	
		\fmffixed{(.3w,0)}{x,w}
		\fmffixed{(.3w,0)}{y,z}
		\fmffixed{(0,-.3h)}{x,y}
		\fmf{photon,tension=0.5}{x,a,w}
		 \fmfv{decoration.shape=circle, decoration.size=.1h, decoration.filled=shaded}{a}
		  \fmfv{decoration.shape=circle, decoration.size=.1h, decoration.filled=shaded}{b}
		\fmf{photon,tension=1}{y,b,z}
                \fmf{electron,  tension=.01}{i1,x,y,i1}
                \fmf{electron,  tension=.01}{o1,z,w,o2}
		 \fmfv{decoration.shape=circle, decoration.size=.05h}{i1}
		 \fmflabel{$j_s$}{i1}
\end{fmfgraph*}
\end{fmffile}
& 
\begin{fmffile}{Additional2}
\begin{fmfgraph*}(120,80)
        \fmfleft{ii1}
        \fmfright{oo1,oo2}
        \fmffixed{(.05w,0)}{ii1,i1}
        \fmffixed{(.05w,0)}{o1,oo1}
        \fmffixed{(.05w,0)}{o2,oo2}
        \fmffixed{(0,.5h)}{o1,v}
        \fmffixed{(0,.5h)}{v,o2}    
        \fmffixed{(.3w,0)}{x,w}
        \fmffixed{(.3w,0)}{y,z}
        \fmffixed{(0,-.3h)}{x,y}
        \fmf{photon,tension=1}{x,a,w}
         \fmfv{decoration.shape=circle, decoration.size=.1h, decoration.filled=shaded}{a}
          \fmfv{decoration.shape=circle, decoration.size=.1h, decoration.filled=shaded}{b}
        \fmf{photon,tension=1}{y,b,z}
                \fmf{electron,  tension=.01}{i1,y,x,i1}
                \fmf{electron,  tension=.01}{o1,z,w,o2}
   		 \fmfv{decoration.shape=circle, decoration.size=.05h}{i1}
		 \fmflabel{$j_s$}{i1}
\end{fmfgraph*}
\end{fmffile}
\end{tabular}
\caption{
Additional contributions to the two-point function.}
\label{additional}
\end{figure}
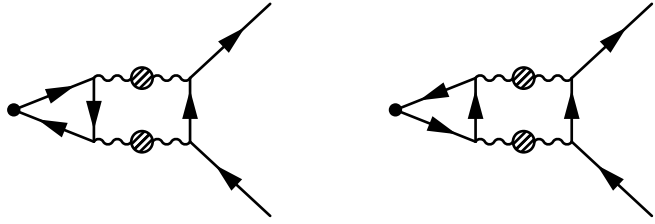

The sum of the two "three point-function" diagrams is given by the following integral, when $s$ is even,
\begin{equation}
	I_3 =-\frac{1}{2}\frac{MN}{2}\tr \left( \gamma^- \int \frac{d^3 q}{(2\pi)^3} \gamma^\mu \frac{1}{i(\slashed{p}-\slashed{q})} \gamma^\nu G_{\mu\alpha}(q) G_{\beta \nu}(q) C^{\alpha \beta}(q)\right)
\end{equation}
where
\begin{equation}
	C^{\mu \nu}(q) = \int \frac{d^3p}{(2\pi)^3} \Tr \left( \frac{1}{i (\slashed{p}-\slashed{q})} \gamma^\mu \frac{1}{i \slashed{p}} V_s \gamma_- p_-^{s-1}  \frac{1}{i \slashed{p}} \gamma^\nu \right) 
\end{equation}

The two diagrams cancel if $s$ is odd. Evaluating the diagrams for even $s$, we obtain:
\begin{equation}
    I_3 = 32 \lambda_M ^2 
   p_-^{s-1} \frac{M}{N}\frac{ \left(\pi ^2 \lambda_M ^2 s^3 +256 s^2  +5 \pi ^2 \lambda_M ^2 s +128
   \right)}{\left(\pi ^2 \lambda_M ^2+64\right)^2
   \left(4 s^4-5
   s^2+1\right)} \log
   \Lambda.
\end{equation}

The contribution to the anomalous dimension is read off from the divergence via:
\begin{equation}
I_3 = V_s p_-^{s-1} \delta_3 (-\log \Lambda).
\end{equation}

From here,
\begin{equation}
    \delta^{(3)}_s=-32 \lambda_M ^2 
  \frac{M}{N}\frac{ \left(\pi ^2 \lambda_M ^2 s^3 +256 s^2  +5 \pi ^2 \lambda_M ^2 s +128
   \right)}{\left(\pi ^2 \lambda_M ^2+64\right)^2
   \left(4 s^4-5
   s^2+1\right)}.
\end{equation}

The anomalous dimension for even $s$ is:
\begin{equation}\begin{split}
    \delta_{\text{even}} & =\delta^{(1)}_s+\delta^{(2)}_s+\delta^{(3)}_s = \frac{M}{N} \frac{8 \lambda_M ^2}{{3 \left(\pi
   ^2 \lambda_M ^2+64\right)^2}} \\ & {\left(6
   \left(\pi ^2 \lambda_M
   ^2+64\right) g(s)+\frac{-256
   \left(11 s^4-s^2+8\right)-4
   \pi ^2 \lambda_M ^2(11 s^4+3 s^3-13 s^2+15 s+2)}{4 s^4-5
   s^2+1}\right)}.
   \end{split} \label{even}
\end{equation}
This is always positive.  One can observe that it vanishes for $s=2$.  

For odd $s$, the third diagram does not contribute and the anomalous dimension is:
\begin{equation}
    \delta_{\text{odd}} = \frac{16 \lambda_M ^2 M \left(3 \left(4 s^2-1\right) H_{s-\frac{1}{2}}+2 s^2 (6 \log
   (4)-11)+4-3 \log (4)\right)}{3 \left(\pi ^2 \lambda_M ^2+64\right) N \left(4
   s^2-1\right)} \label{odd}
\end{equation}
One can observe that it vanishes for $s=1$.

For large spin, we have:
\begin{equation}
  \delta_s \sim \frac{M}{N} \frac{16 \lambda_M ^2}{\left(\pi ^2 \lambda_M
   ^2+64\right)} \log s
\end{equation}

\begin{figure}[h!]
\begin{center}
\includegraphics[width=12cm]{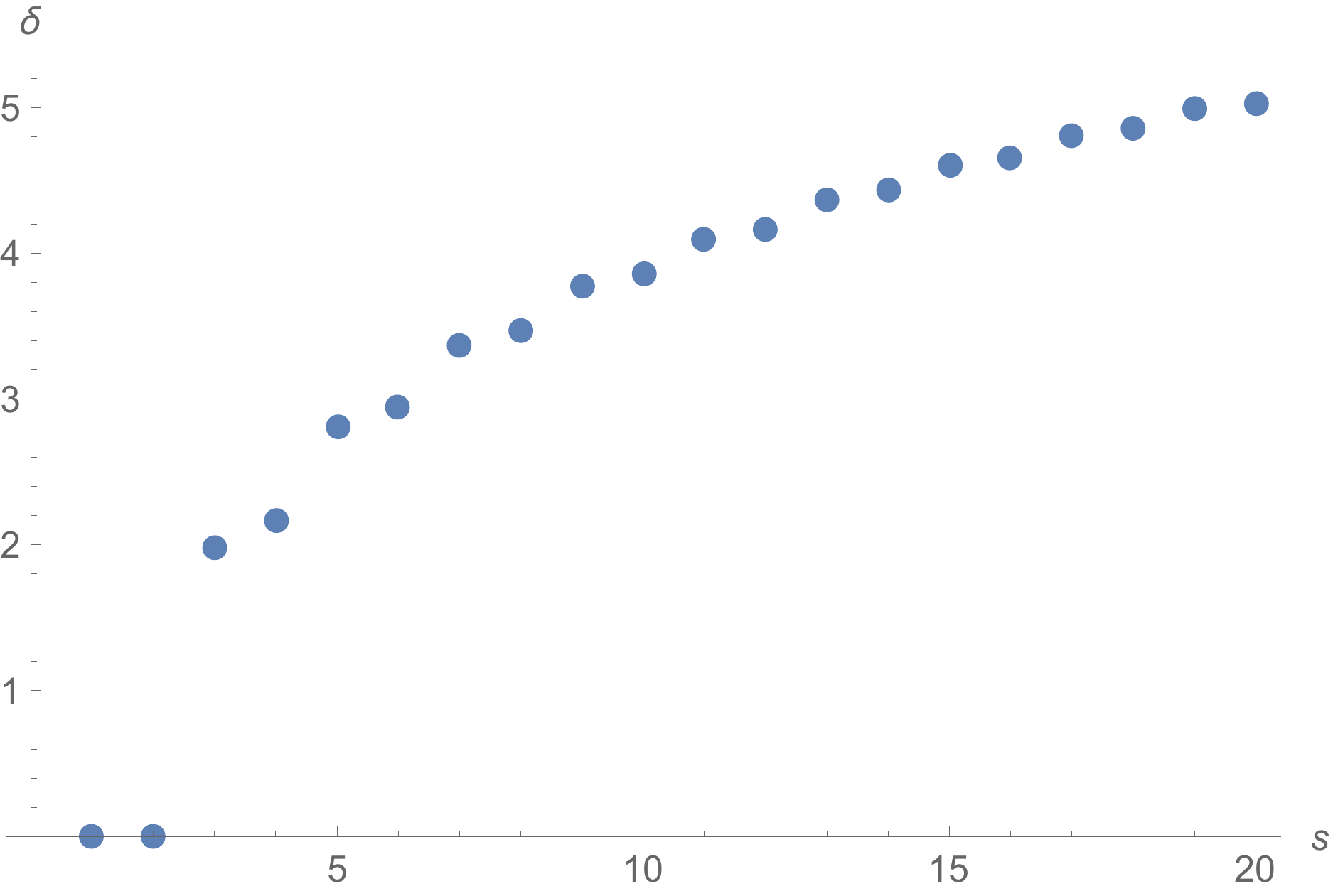}
\caption{ Anomalous dimensions of the current as a function of spin, in the strong coupling limit, (to be multiplied by $M/N$). \label{spin-plot}}
\end{center}
\end{figure}

At large coupling, for even spin, we have
\begin{equation}
    \delta=\frac{M}{N} \frac{16}{3 \pi ^2} \left(3
   H_{s-\frac{1}{2}}-\frac{2 \left(11 s^4+3 s^3-13 s^2+15
   s+2\right)}{4
   s^4-5 s^2+1}+\log
   64\right).
\end{equation}
where, $H_{s-\frac{1}{2}}$ is a harmonic number. At large coupling the anomalous dimensions approach a constant value, which increases logarithmically with $s$. The strong coupling limit for both even and odd spin is plotted in Figure \ref{spin-plot}.

\begin{figure}
\begin{center}
\includegraphics[width=12cm]{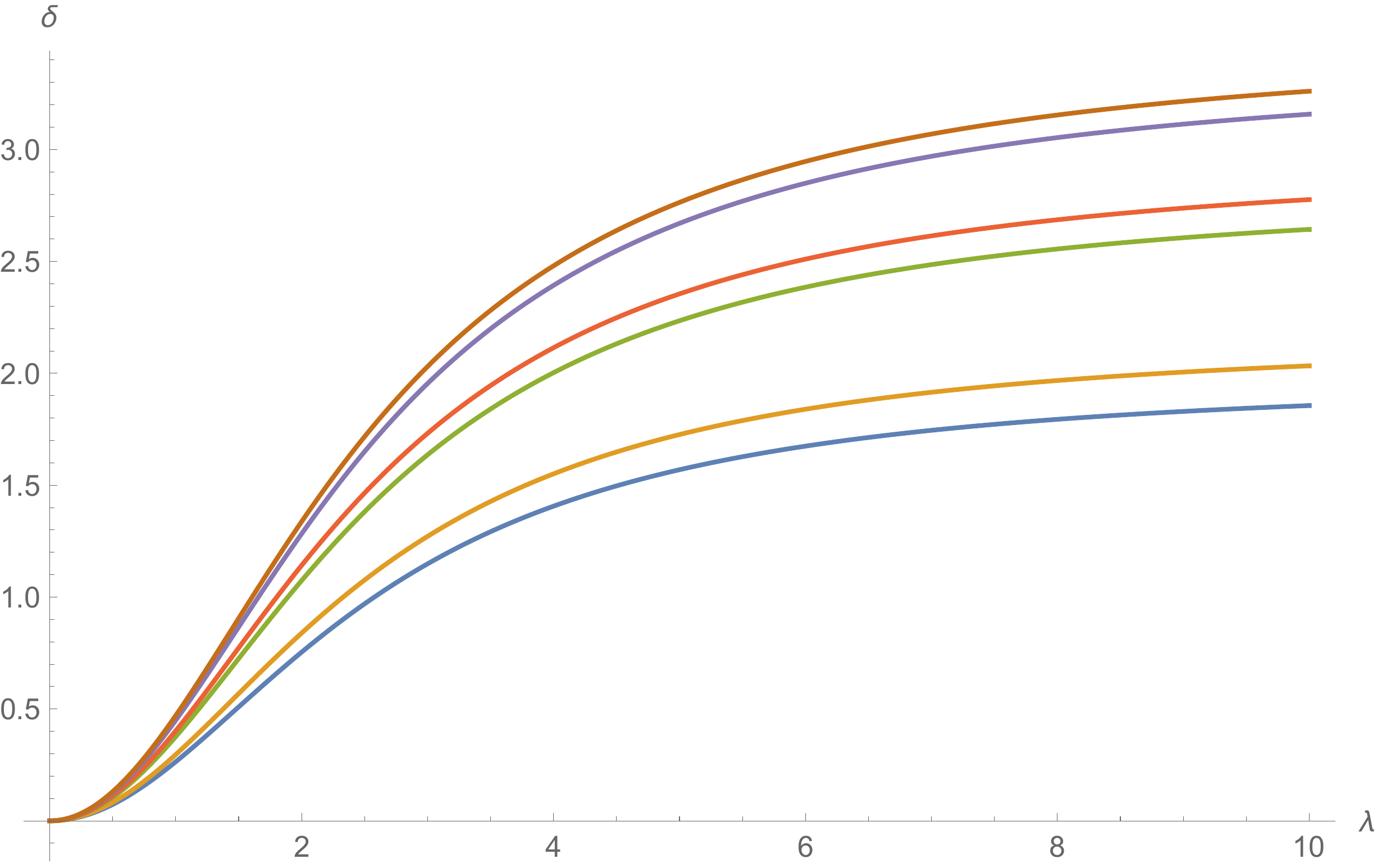}
\caption{Anomalous dimension of the spin $s$ current, for $s=3,4,5,6,7,8$ as a function of $\lambda_M$ (to be multiplied by $M/N$).}
\end{center}
\end{figure}

\section{Large Spin Expansion}
\label{sec-3}
In this section, we briefly comment on the large spin expansion of our results and the results of \cite{Giombi:2016zwa}. 

In the large spin limit, our results can be written as:
\begin{equation}
\begin{split}
    \delta_s=& \frac{M}{N} \left(\frac{8\lambda_M ^2}{3 \pi ^2 \lambda_M ^2+192}\right)
     (6 \log (4 s)+6 \gamma -11) \\
     & +
    \frac{M}{N}\left(\frac{\lambda_M ^2}{3 \pi ^2 \lambda_M ^2+192}\right)
    \left(-\frac{4}{s^2}-\frac{37}{20 s^4}-\frac{4}{21
   s^6}+O\left(\frac{1}{s^7}\right)\right), \\
   & \text{ for odd $s$,}
   \end{split}
   \label{odd-large-s}
\end{equation}
and
\begin{equation}
\begin{split}
    \delta_s= & \frac{M}{N}\left(\frac{8\lambda_M ^2}{3 \pi ^2 \lambda_M ^2+192}\right)
     (6 \log (4 s)+6 \gamma -11) \\
    & +
   \frac{M}{N} \left(\frac{512 \lambda_M ^2}{3 \left(\pi ^2 \lambda_M ^2+64\right)^2} \right)
    \left(-\frac{25}{2
   s^2}-\frac{3397}{160 s^4}-\frac{1955}{84 s^6}+O\left(\frac{1}{s^7}\right) \right) \\ &+
   \frac{M}{N}\left(\frac{8 \pi ^2 \lambda_M ^4}{3 \left(\pi ^2 \lambda_M ^2+64\right)^2}\right)\left(-\frac{3}{s}-\frac{1}{2 s^2}-\frac{75}{4 s^3}-\frac{37}{160
   s^4}+O\left(\frac{1}{s^5}\right)\right),\\
   & \text{ for even $s$.}
   \end{split} \label{even-large-s}
\end{equation}

We presented the anomalous dimension in this way to draw an analogy with equations (61) and (63) of \cite{Alday:2015eya}, which give the large spin expansion of the large $N$ anomalous dimensions of non-singlet and singlet operators in the critical $O(N)$ model \cite{Lang:1992zw, Lang:1993ge}.

The first term in each equation is the $s\rightarrow \infty$ limit of the anomalous dimension, which contains a $\log s$, unlike the critical $O(N)$ model. Hence the anomalous dimension of $\psi$ is undefined, which is expected as it is not a gauge invariant operator. 

Looking at the subsequent terms in equations \eqref{odd-large-s} and \eqref{even-large-s}, it appears as if, just as in the critical $O(N)$ model, for odd-spins (which resemble non-singlet operators in the critical $O(N)$ model), the anomalous dimensions are due to an intermediate operator of twist two (i.e., $\bar{\psi}\psi$),
while for even spins (which resemble singlet operators in the critical $O(N)$ model) we have two series -- the first series corresponding to the twist-two operator and the second series corresponding to twist-one operators (such as $j_s$). There may be some ambiguity in separating these two series, and these observations are only suggestive at present because the  the analysis of \cite{Alday:2015eya} does not immediately apply here.

\subsection{Conformal Spin}
It is also interesting to express these results in terms of the conformal spin $J$ \cite{Alday:2015eya}, defined as a function of spin $s$ and twist $\tau=\Delta-s$ as:
\begin{equation}
    J^2=(s+\tau/2-1)(s+\tau/2).
\end{equation}

In \cite{Alday:2015eya}, it was argued that the expansion of anomalous dimensions around large conformal spin should contain only even powers of $1/J$, possibly with extra $\log J$ coefficients. Due to the fact that the field $\psi$ is not a gauge invariant operator, and the dimensionality of space-time is odd, the results of \cite{Alday:2015eya} do not immediately apply here. However, as will be described elsewhere, applying the ideas of \cite{ Alday:2015ota, Alday:2016njk, Alday:2016jfr}, it is possible to refine the arguments of \cite{Alday:2015eya} to show that this result holds to order $\lambda_M^2$, but should not be expected to hold at higher orders.\footnote{We thank Luis Fernando Alday for correspondence on this point.}

Expanded in powers of $M/N$, the conformal spin is given by 
\begin{equation}
    J=J_0+J_1\frac{M}{N}+ \ldots.
\end{equation}
Because the leading order anomalous dimensions vanish in our theory, $\tau=1+O(\frac{M}{N})$ and $J_0$ is given by the free value:
\begin{equation}
    J_0^2=(s+1/2)(s-1/2).
\end{equation}

Re-expressing our results in terms of $J=J_0+O\left(\frac{M}{N}\right)$, we have, for odd spins:
\begin{equation}
\begin{split}
    \delta^{\text{odd}}_s(J) =&\frac{M}{N} \Bigg[ \frac{8 \lambda_M ^2 (-11+6 \gamma +12 \log (2))}{3 \left(\pi ^2 \lambda_M ^2+64\right)}+\frac{16\lambda_M^2}{64+\pi^2\lambda_M^2} \psi\left(\sqrt{ J^2+\frac{1}{4}}+\frac{1}{2}\right)
    \\ &- \frac{2 \lambda_M ^2}{ 
   \left(\pi ^2 \lambda_M ^2+64\right)} \frac{1}{J^2}\Bigg],
    \end{split}
\end{equation}
Here $\psi$ denotes the digamma function. Note that 
\begin{equation}
   \psi\left( \sqrt{ J^2+\frac{1}{4}}+\frac{1}{2}\right) = \log \left({J}\right)+\frac{1}{6 J^2}-\frac{1}{30
   J^4}+\frac{4}{315 J^6}-\frac{1}{105 J^8}+\ldots 
\end{equation}
contains only even powers of $1/J$ in its expansion, apart from the initial $\log J$. Hence the entire expression has an expansion in even powers of $1/J$, to all orders in $\lambda_M$. 

For even spins, 
\begin{equation}
\begin{split}
    \delta_s(J) =&\frac{M}{N} \Bigg[ \frac{8 \lambda_M ^2 (-11+6 \gamma +12 \log (2))}{3 \left(\pi ^2 \lambda_M ^2+64\right)}+\frac{16\lambda_M^2}{64+\pi^2\lambda_M^2} \psi\left(\sqrt{ J^2+\frac{1}{4}}+\frac{1}{2}\right)
    \\ &
    - \frac{64\lambda_M^2}{\left(\pi ^2 \lambda_M ^2+64\right)^2}
    \left[\frac{2  \left(68 J^2
    +45\right)}{J^2
   (4 J^2-3)}\right]
   \\ 
   & - \frac{2\pi ^2 \lambda_M^4}{\left(\pi ^2 \lambda_M ^2+64\right)^2}
   \left[\frac{ (8J^2 +42) \sqrt{4
   J^2+1}}{
   J^2 (4 J^2-3)}+\frac{1}{J^2}
   \right]\Bigg].
    \end{split}
\end{equation}
Here, we see that the last piece would contain even as well as odd powers of $1/J$. This corresponds to an order $\lambda_M^4$ correction, so at order $\lambda_M^2$, we again get an expansion with only even powers of $1/J$. The odd powers of $1/J$  originate from a term proportional to $\sqrt{4J^2+1}$, and could be interpreted as originating from the exchange of a tower of twist-one (almost-conserved) operators. Such a term is also present in the singlet large $N$ anomalous dimensions of the critical $O(N)$ model in $d=3$, given in equation (64) of \cite{Alday:2015eya}. 

\section{Discussion}
In this paper, we studied the $M/N$ anomalous dimensions in a bifundamental $U(N)\times U(M)$ Chern-Simons theory, $\delta_s(\lambda_M,\lambda_N)$, in the special case $\lambda_N=0$. These results compliment the results of \cite{Giombi:2016zwa}, which effectively determine the anomalous dimensions in the case $\lambda_M=0$.  It should also be possible to determine the $M/N$ anomalous dimensions in the bifundamental theory to all orders in both $\lambda_M$ and $\lambda_N$ by extending the arguments given in section \ref{hs}, however, the calculation would be substantially more involved. We hope to address this problem in the near future. 

Chern-Simons theories coupled to matter have also played an important role in the study of fractional quantum Hall transitions, e.g., \cite{WenWu, Mott, Hui:2017pwe}. In particular, \cite{Mott} includes a study of $U(1)$-Chern Simons theory coupled to $N$ flavours of fermions as a model of a Mott Insulator -quantum Hall transition, which, at the level of $1/N$ corrections also has the same spectrum as the $U(M)$ theory we consider. The anomalous dimension of the scalar operator $\bar{\psi}\psi$ effectively determines the critical exponent $\nu$ associated with the transition, and so is of experimental interest. To date, the anomalous dimension of $\bar{\psi}\psi$ in the large colour theory is not known, and it would be interesting to compare the value of $\nu$ from this calculation to the existing large flavour calculation.

We are not aware if the anomalous dimension of the higher-spin currents have any similar physical interpretation or relevance to Quantum Hall physics, but it may be interesting to pursue this further. It might also be possible to study the spectrum at order $1/N^2$, which may perhaps provide better agreement with experiment (in which $N=1$). The techniques involved would be similar to the recent computations of $1/N^2$ spectrum of the critical $O(N)$ and Gross-Neveu models \cite{Manashov:2016uam, Manashov:2017xtt}. Correlation functions involving the stress tensor and conserved $U(1)$ current could also be calculated in this theory possibly at finite temperature/chemical potential (e.g., \cite{Geracie:2015drf}), and may correspond to physical observables. These should obey the bounds determined in \cite{Hofman:2008ar,Chowdhury:2016hjy, Chowdhury:2017vel, Cordova:2017zej}. 

It would also be of interest to consider scaling dimensions in the theory of $U(M)_k$ Chern-Simons coupled to $N$ critical bosons, when $N$ is large. In \cite{WenWu}, the anomalous dimension of the scalar operator $j_0$ was calculated in the $U(1)$ version of this theory. It would be interesting to also compute the higher spin spectrum and compare it is to the spectrum of the fermionic theory studied here. 

Chern-Simons theories coupled to matter also exhibit a beautiful bosonization duality \cite{Aharony:2012nh} relating theories with fermionic matter to critical bosonic matter. This duality was discovered and has been supported via various large $N$ computations \cite{MZ, GurAri:2012is, Aharony:2012ns, Jain:2013py, Jain:2013gza, Jain:2014nza, Bedhotiya:2015uga, Gur-Ari:2015pca,Yokoyama:2016sbx,Minwalla:2015sca,Gur-Ari:2015pca}, but it is now conjectured to hold at finite $N$ as well \cite{Aharony:2015mjs, Kachru:2016rui, Kachru:2016aon, Hsin:2016blu,Aharony:2016jvv}, and has been of interest from the point of view of condensed matter (e.g., \cite{Seiberg:2016gmd,Karch:2016sxi, Murugan:2016zal, Metlitski:2016dht}). These dualities can be thought of as generalizations of the supersymmetric Giveon-Kutasov duality \cite{Giveon:2008zn, Kapustin:2011gh, Benini:2011mf,  Intriligator:2013lca}. While these dualities cannot hold in the large flavour limit, they may perhaps be generalized to bifundamental Chern-Simons theories when both couplings are nonzero, as a strong/weak duality when $\lambda_N \leftrightarrow (1-|\lambda_N|)$. One form for these dualites was conjectured in \cite{Gurucharan:2014cva}, based on analogy with ABJ dualities \cite{Aharony:2008gk}, and could be tested by explicit computation when $M\ll N$ in the fermionic and critical bosonic theory. (Some perturbative calculations in the non-critical bifundamental bosonic theory appear in \cite{Banerjee:2013nca}.) We hope to perform these tests in the near future. 

Finally, we found the spectrum of anomalous dimensions appears to take a form similar to those of the large $N$ critical $O(N)$ model, when expanded in a large spin expansion, and also expressed in terms of conformal spin, despite the additional logarithmic dependence on spin. It would be interesting if this behaviour could be better understood from the point of view of large spin perturbation theory, e.g., \cite{Alday:2015eya, Alday:2015ota, Alday:2016njk, Alday:2016jfr}.

\section*{Acknowledgements}

The authors would like to thank Luis Fernando Alday for extensive discussions and reading through a copy of this draft. The authors would also like to thank Eugene Skvortsov and Shiraz Minwalla for discussions. This work is partially supported by a DST INSPIRE Faculty Award.

\bibliographystyle{JHEP}
\bibliography{CSBib}
\end{document}